\documentclass{article}
\usepackage[dvips]{graphicx}
\begin{document}

\begin{center}
Talk given at the 2K1BC Workshop\\
 {\it Experimental Cosmology at Millimetre Wavelengths}\\
 July 9-12, 2001, Breuil-Cervinia (Italy)
\end{center}

\vskip2mm
\hrule
\vskip1cm
\begin{center}
\huge{\bf Big Bang Nucleosynthesis, Cosmic Microwave Background 
Anisotropies and Dark Energy}
\vskip4mm
\large{\bf M. Signore$^\ast$, D. Puy$^\dagger$}
\vskip4mm
{\small
$^\ast$Observatoire de Paris-DEMIRM, Paris (France), 
email: monique.signore@obspm.fr
\vskip1.5mm
$^\dagger$Institute of Theoretical Physics, Z\"urich and PSI-Villigen 
(Switzerland), email: puy@physik.unizh.ch}
\end{center}
\vskip9mm
\noindent
{\bf Abstract.}
Over the last decade, cosmological observations have attained a level of 
precision which allows for very detailed comparison with theoretical 
predictions. We are beginning to learn the answers to some fundamental 
questions, using information contained in Cosmic Microwave Background 
Anisotropy (CMBA) data.  
\\
In this talk, we briefly review some studies of the current and 
prospected constraints imposed by CMBA measurements on the neutrino physics 
and on the dark energy. As it was already announced by Scott 
\cite{scott:1999}, we present some possible {\it new physics} from the 
Cosmic Microwave Background (CMB).
\section{Introduction}
Since the 80's, cosmologists introduce for the baryonic density 
($\rho_b$ or $\Omega_b$) of the Universe, a concordance interval where 
predicted and 
measured abundances of light elements ($^7Li$, $^4He$, $D$) were consistent, 
within their uncertainties. At the end of 90's, the determination of primeval 
$D$ is supposed to be accurate enough to pin down $\rho_b$ or $\Omega_b$; 
and the {\it concordance intervals} for $D$, $^7Li$, $^4He$ and $\Omega_b$ 
were predicted by $D$ measurements, see Burles et al. \cite{burles:1999} and 
Signore-Puy \cite{signore:1999}. Fig. (1) shows this problem of concordance 
from the works of Burles-Nollett-Turner \cite{burles:1999}.

\begin{figure}[h]
\begin{center}
\resizebox{.4\columnwidth}{!}
  {\includegraphics[angle=0]{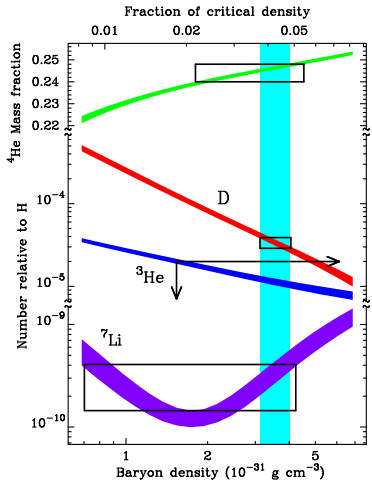}}
\caption{The situation of BBN in 1999. It shows the concordance intervals 
for each element ($2\sigma$ uncertainties) and the baryon density predicted 
by primordial deuterium measurements, from Burles-Nollett-Turner 
\cite{burles:1999}.}
\end{center}
\end{figure}
\noindent
In the year 2000, observations of CMBA have become a competitive means for 
estimating $\Omega_b$. Then CMBA results can be a cross check of Big Bang 
nucleosynthesis (BBN) results and can lead also to new constraints on BBN 
theory, ie neutrino physics.
\\
On the other hand from observations of type Ia-supernova at high $z$ carried 
out by two major teams -{\it Supernovae Cosmology Project} 
\cite{perlmutter:1999} and {\it High-z Supernovae Team} \cite{riess:1998}- 
there is some direct evidence that the present Universe is accelerating. For 
a pedagogical and recent review on this subject see \cite{signore:2001} and 
references therein. Moreover, recent measurements of CMBA and of baryon 
fraction in galaxy clusters indicate that the Universe is flat and that the 
matter contributes about one third of the critical density $\Omega_M \sim 
1/3$; and about two thirds of the critical density constitutes the 
{\it dark energy}: $\Omega_\Lambda \sim 2/3$. The nature of this dark energy 
is the new challenge for cosmology and fundamental physics. Now, an important 
question is: besides the {\it SNIa experiments} (see SNAP \cite{SNAP:2001}), 
can CMBA measurements provide constraints on the nature of the dark energy ?

\section{Constraints on Neutrino Physics}

As already said above, until recently, BBN -from observations of primordial 
$D$ abundances- provided the only precision estimates of $\Omega_b$. 
Considering the most recent primordial deuterium data, Burles, Nollett 
and Turner \cite{burles:2001} give:
\begin{equation}
\Omega_b \, h^2 \, = \, 0.017 - 0.024 \ \ (95 \% \ {\rm CL})
\end{equation}
where $h$ is the Hubble parameter in units of 100 km sec$^{-1}$ Mpc$^{-1}$. 
In the past year, the first results which may rightly be called precision 
CMBA measurements have been obtained from BOOMERANG \cite{debernardis:2000} 
and MAXIMA \cite{hanany:2000}. A higher baryon density than that predicted from BBN has been claimed:
\begin{equation}
\Omega_b \, h^2 \, \sim \, 0.03 .
\end{equation}
The discrepancy between BBN and CMBA estimates for $\Omega_b$ led to the 
suggestions that one must consider some {\it new physics} which appeared 
between the BBN epoch ($T \sim 1$ MeV) and the CMB epoch ($T \sim 1$ eV) 
in order to understand these different values of $\Omega_b$ (
\cite{lesgourgues:2000}, \cite{hannestad:2000}, \cite{orito:2000}, 
\cite{esposito:2000}, \cite{mangano:2001}).
\\
But the more recent data from BOOMERANG \cite{netterfield:2001}, 
MAXIMA \cite{lee:2001} and DASI \cite{halverson:2001} show that 
there is no more difference between BBN and CMBA estimates for $\Omega_b$:
\begin{equation}
\Omega_b \, h^2 \, = \, 0.02,
\end{equation}
and therefore no need of {\it new physics} to reconcile BBN and CMBA data. 
However, some cosmologists, in particular 
Kneller et al. \cite{kneller:2001}, Hannestad \cite{hannestad:2001}, Hansen et 
al. \cite{hansen:2001} consider these CMBA data sets of high precision to 
constrain, independently of BBN, this {\it new physics}, that is to say 
the {\it neutrino physics}.
\subsection{BBN limit on $N_\nu$}
First, let us introduce $N_\nu$ - the equivalent number of standard model 
neutrino species- through the energy density $\rho$:
\begin{equation}
N_\nu \, \equiv \, \frac{\rho}{\rho_{\nu_o}}
\end{equation}
where $\rho_{\nu_o}$ is the energy density of a standard neutrino species. 
This is a way of expressing the energy density in light non-interacting 
species. As noted in \cite{hannestad:2001}, the standard model 
predicts:
\begin{equation}
N_\nu \, \sim \, 3.04
\end{equation}
due to the fact that the neutrinos are not completely decoupled during 
the $e^+-e^-$ annihilation; see Steigman \cite{steigman:2001} for a detailed 
neutrino counting and a discussion on the above value, Eq. (5). 
The abundances of primordial $^4He$, $D$, $^7Li$ can be used to determine 
BBN limit on $N_\nu$. For example, Lisi et al. \cite{lisi:1999} give the 
following bound adopted also by \cite{hannestad:2001}:
\begin{equation}
3 \, \leq \, N_{\nu, BBN} \, \leq \, 4 \ \ (95 \% \ {\rm CL}) .
\end{equation}
Let us also mention the work done by Kneller et al. \cite{kneller:2001} who 
consider for BBN predictions the three parameters $\eta$ -the baryon to photon 
ratio ($\eta_{10} = 10^{10} \times \eta = 274 \, \Omega_b h^2$ - 
$\Delta N_\nu$ the asymetrical part (or degenerate part) of $N_\nu$ and 
$\xi_\nu \equiv \mu_\nu / T_\nu$ where $\mu_\nu$ and $T_\nu$ are 
respectively the chemical potential and the temperature of the $\nu$-species.
\subsection{CMBA limit on $N_\nu$}
A bound on $N_\nu$ has also been derived from CMBA data by many authors 
\cite{kneller:2001} \cite{hannestad:2001} \cite{hansen:2001}. 
Let us only summarize the main points of these studies:
\begin{itemize}
\item {\it i)} Kneller et al. \cite{kneller:2001} used the CMFAST software 
\cite{seljak:1996} in order to calculate the cosmic background fluctuation 
spectrum as a function of $\eta$ and $\Delta N_\nu$ and compare to 
BOOMERANG \cite{netterfield:2001}, MAXIMA \cite{lee:2001} and DASI 
\cite{halverson:2001} observations for four different cosmological 
models. They show, in the ($\eta-\Delta N_\nu$) plane, the four very different 
shapes of the confidence interval contours corresponding to the four 
cosmological models. Their results point out the sensitivity of the 
{\it new physics} ($\Delta N_\nu$) to the other cosmological parameters. 
\item {\it ii)} The analysis of the CMBA data by three groups 
\cite{kneller:2001}, \cite{hannestad:2001}, \cite{hansen:2001} lead to robust 
upper bounds on $N_\nu$ 
\begin{equation}
N_\nu \, < \, 7-17
\end{equation}
which are much weaker than that given from BBN data, the right hand side 
of equation (6) ! 
\item {\it iii)} Adding large scale structure data to CMBA data Hannestad 
\cite{hannestad:2001} gives a non trivial lower bound:
\begin{equation}
N_\nu \, > \, 1.5 \ \ (95 \% \ {\rm CL})
\end{equation}
which is the first independent indication of the presence of a cosmological 
neutrino background, predicted by the standard model, and already seen in 
BBN data, the left hand side of equation (6).
\item {\it iV)} It seems that there is no significant indication of non 
standard physics -i.e. no {\it new physics}- contributing to $N_\nu$ at the 
recombination epoch \cite{kneller:2001} \cite{hannestad:2001}.
\end{itemize}

\section{Constraints on dark energy}

Recent observations \cite{perlmutter:1999} \cite{riess:1998} of type 
Ia-supernovae indicate that the Universe may be presently dominated by an 
additional {\it dark energy} with a negative pressure such that the Universe 
is presently accelerating (see also \cite{signore:2001}). Combined observations 
of type Ia-supernovae \cite{perlmutter:1999}, CMBA \cite{jaffe:2000} 
and cluster evolution \cite{bahcall:1998} for 
which the results have been done in the form of likelihood contours in the 
$\Omega_M$ and $\Omega_\Lambda$ plane are reported in Fig. (2). 
\\
$\Omega_M$ and $\Omega_\Lambda$ are defined by 
\begin{equation}
\Omega_M \, = \, \frac{8 \pi G \rho_o}{3 H_o} \ , \ \ 
\Omega_\Lambda  \, = \, \frac{\Lambda}{ H_o}
\end{equation}
where the index $o$ refers to the present epoch, $\rho$, $H$ and $\Lambda$ 
being respectively, the energy density, the Hubble parameter and the cosmological 
constant -see \cite{signore:2001} for instance. 

\begin{figure}[h]
\begin{center}
\resizebox{.6\columnwidth}{!}
  {\includegraphics{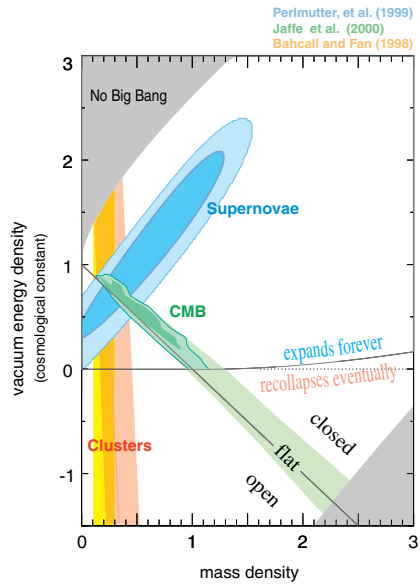}}
\caption{From \cite{SNAP:2001}, Confidence regions in the 
($\Omega_M - \Omega_\Lambda$) plane for high z-supernovae 
(see Perlmutter et al. \cite{perlmutter:1999}), CMBA \cite{jaffe:2000}) and 
cluster evolution (Bahcall \& Fan \cite{bahcall:1998}) measurements. The 
consistent {\it overlap} is a strong indicator for the existence of 
$\Omega_\Lambda$, i.e. of a cosmology constant or dark energy.}
\end{center}
\end{figure}

Let us recall that the Friedman-Lemaitre equation can be written as
\begin{equation}
\Omega_M + \Omega_\Lambda + \Omega_k =1 
\end{equation}
with a term of matter plus radiation $\Omega_M$, a term of {\it dark energy} 
$\Omega_\Lambda$ and a curvature term such that 
\begin{equation}
\Omega_k \, = \, \frac{k}{R_o^2 H_o^2}
\end{equation}
where $R$ is the cosmic scale factor and $k$ is the curvature constant.
\\
{\it What is the nature of this dark energy ?} This is the present challenge 
for cosmology and particle physics. The simplest  interpretation of this dark 
energy is the {\it cosmological constant $\Lambda$} (vacuum energy) for which 
the equation of state:
\begin{equation}
w \, \equiv \, \frac{P}{\rho}
\end{equation}
is equal to $-1$.
\\
It is important to know if this cosmological {\it constant}, as inferred by 
observations, is truly constant or if the observations point out some form 
of cosmic evolution often called {\it quintessence} $Q$, for which the equation 
of state $w_Q$ is such that:
\begin{equation}
-1 \, \leq \, w_Q \, \leq \, 0 .
\end{equation}
In this case, the vacuum energy is the result of a scalar field $Q$ slowly 
evolving along an effective potential or getting trapped in a local minimum and 
which only interacts with the other fields via gravity- In any case 
-cosmological constant or quintessence- one is faced with two problems:
\begin{itemize}
\item $i)$ a fine tuning problem: why the vacuum energy, is so small ? From 
particle physics, one might expect: 
$\Lambda/8\pi G \sim m^4_{planck}$, and it is 
off by about 120 orders of magnitude.
\item $ii)$ a cosmic coincidence problem: why $\Omega_M$ and $\Omega_\Lambda$ are 
nearly equivalent now ?
\end{itemize}
Since $w$ is, in general, time vrying, the first step toward solving the 
dark energy problem is to determine $w(t)$ or $w(z)$.
\\
Before considering some models of quintessence, let us only recall that:
\begin{itemize}
\item $\bullet$ for the case where the dark energy is the cosmological 
constant $\Lambda$:
\begin{equation}
w_\Lambda \, = \, -1 
\end{equation}
\item $\bullet$ some authors -for instance, Huey et al. \cite{huey:1999}- 
introduce an effective (constant) equation of state $w_{eff}$ defined 
by:
\begin{equation}
w_{eff} \, \sim \, \frac{\int \, \Omega_Q(z) \, \omega(z) \, dz }
{\int \, \Omega_Q (z) \, dz}
\end{equation}
\item $\bullet$ for topological defects:
\begin{equation}
w_{string} \, \sim \, -\frac{1}{3} \ \ {\rm and}  \ \ 
w_{wall} \, \sim \, -\frac{2}{3}
\nonumber
\end{equation}
\end{itemize}
\subsection{On Quintessence Models}
Many quintessence effective potentials exist in the litterature -see, for 
instance Weller \& Albrecht \cite{weller:2001a}. Here, let us only mention:
\begin{itemize}
\item $\bullet$ \underline{The cosmological tracker solutions} 
\cite{steinhardt:1999} \cite{zlatev:1999}\\
 with, in particular,  the inverse 
tracker potential of Ratra \& Peebles \cite{ratra:1988}
\begin{equation}
V(Q) \, = \, M^{(4+\alpha)} \, Q^{-\alpha}
\end{equation}
where $M$ and $\alpha$ are parameters such as 
$\Omega_Q \sim 2/3$ at present. The tracker solutions evolve on a common 
evolutionary track independent of the initial conditions. 
All the tracker models have in common that the density in the dark energy 
at late times dominates over all the other density contributions 
and therefore the expansion of the Universe starts accelerating.
\item $\bullet$ \underline{The Supergravity Potential}:\\
\begin{equation}
V_{SUGRA} (Q) \, = \, M^{(4+\alpha)} \, Q^{-\alpha} \, 
{\rm exp} \Bigr[\frac{1}{2} \Bigr( \frac{Q}{M_{pl}} \Bigl)^2 \Bigl], 
\end{equation}
which is related to the supersymmetry breaking -see in particular Binetruy 
\cite{binetruy:1999}, Brax \& Martin \cite{brax:1999}. 
$M$ and $\alpha$ are chosen such that the supersymmetry breaking occurs 
above the electroweak scale. A discussion on this potential is found in 
Kolda \& Lyth \cite{kolda:1999}.
\end{itemize}
In all of these models, the energy density of the field $Q$ is given by the 
kinetic and potential components:
\begin{equation}
\rho_Q \, = \, \frac{1}{2} \, \dot{Q}^2 + V(Q)
\end{equation}
while the pressure is given by the difference 
\begin{equation}
P_Q \, = \, \frac{1}{2} \, \dot{Q}^2 - V(Q) .
\end{equation}
Moreover, we assume that the field $Q$ is homogeneous on large scales. 
Therefore the equation of state of the quintessence is given by:
\begin{equation}
w_Q \, = \, \frac{P_Q}{\rho_Q} .
\end{equation}
Fig. (5) in Weller \& Albrecht \cite{weller:2001a} shows the evolution 
-in the range of redshift $[0-2]$- of the equation of state of the dark 
energy component:
\begin{equation}
w_Q \, = \, w_Q(z)
\end{equation}
for all of these models they discuss in \cite{weller:2001a} and, in 
particular, for the two potentials considered here, Eqs (17) and (18).
\subsection{Constraints on the equation of state of dark energy}
We have seen that searches for SNIa at high $z$ have already provided a strong 
evidence for an accelerating present Universe \cite{perlmutter:1999}, 
\cite{riess:1998}, \cite{signore:2001}. By analyzing a simulated data set as 
might be obtained by the proposed SNAP satellite \cite{SNAP:2001}, Weller \& 
Albrecht \cite{weller:2001b} claim that it will be possible to discriminate 
among different dark energy solutions. 
\\
Fig. (3) shows the separation of three dark energy models in the 
($\Omega_M - w_o$) plane where $w_o$ is such that 
$w = w_o + w_1 z$, although Maor et al. 
\cite{maor:2001} show that this method is indeed very limited.  
\\
However, as already seen through the Fig. (2), the result can be better by 
combining SNIa constraints with other complementary measurements. A low-$z$ 
measurement such as a cluster survey , an intermediate-$z$ measurement 
such as a SNIa survey and a high-$z$ measurement such as CMBA 
measurements can provide complementary constraints.

\begin{figure}[h]
\begin{center}
\resizebox{.45\columnwidth}{!}
  {\includegraphics{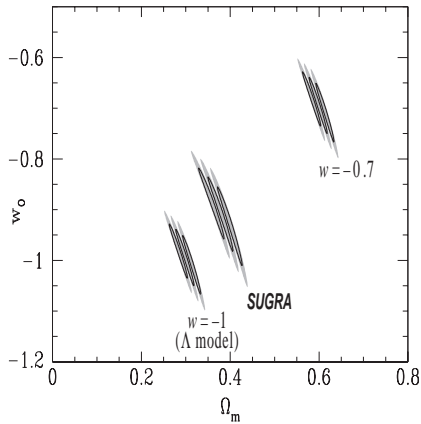}}
\caption{From Weller \& Albrecht \cite{weller:2001b}, separation 
of three dark energy models in the ($\Omega_M-w_o$) plane.}
\end{center}
\end{figure}

Fig. (4) from Hu et al. \cite{hu:1999} and Fig. (5) from 
Huterer \& Turner \cite{huterer:1999} indicate how the constraints from 
several measurements would constrain a model when the Universe is assumed 
flat ($\Omega_k =0$) and $\omega$ is supposed to be constant

\begin{figure}[h]
\begin{center}
\resizebox{.7\columnwidth}{!}
  {\includegraphics{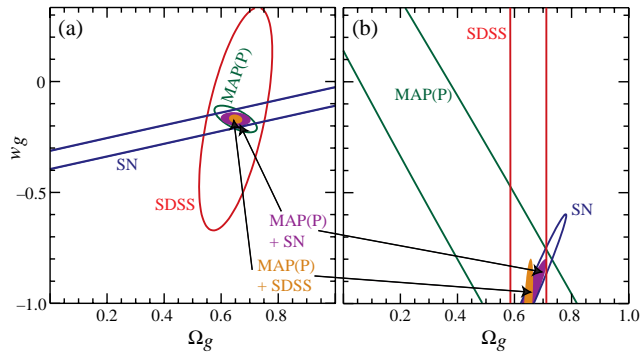}}
\caption{From Hu et al. \cite{hu:1999}, here 
$\Omega_g=\Omega_Q=1-\Omega_M$, confidence regions in the 
$(\Omega_g - w_g)$ plane from CMB, SN and large scale structure survey (68 \% CL). 
Here \underline{SN} means constraints of a supernova program such as 
SNAP \cite{SNAP:2001}, \underline{SDSS} means constraints of Sloan Digital Sky Survey 
\cite{SDSS:2001}, \underline{MAP} means constraints of the MAP satellite \cite{MAP:2001}; 
(P) means polarization information. \texttt{(a)}: Left curves, 
$w_g=w_Q=-1/6$, $\Omega_M \sim 1/3$. \texttt{(b)}: Right curves, 
$w_g=w_Q=-1$, $\Omega_M \sim 1/3$. Note the complementarity nature of the 
data sets. }
\end{center}
\end{figure}

\begin{figure}[h]
\begin{center}
\resizebox{.5\columnwidth}{!}
  {\includegraphics{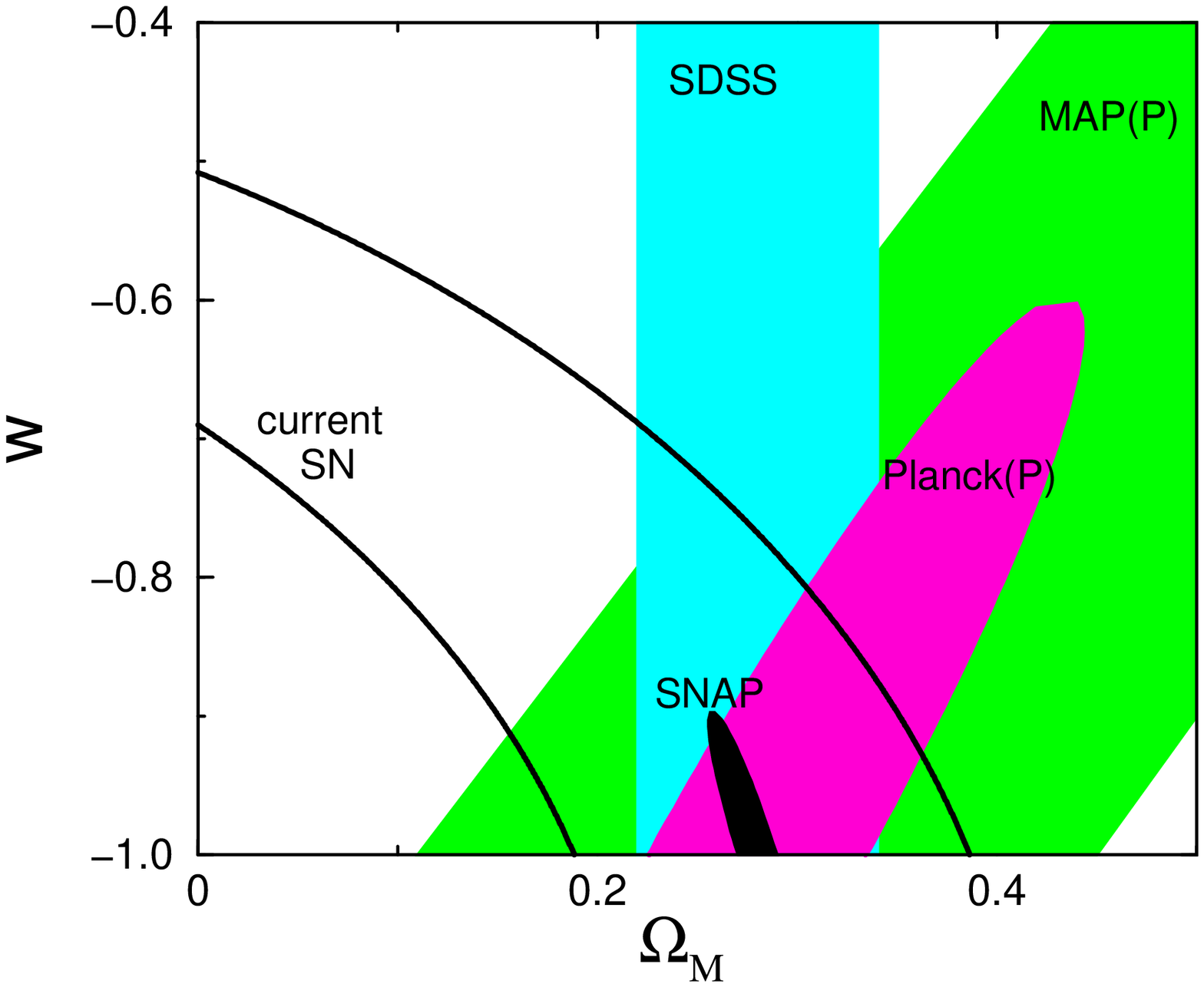}}
\caption{From Huterer \& Turner \cite{huterer:1999}, Confidence regions in 
the $(\Omega_M, w)$ plane for the case b) of Fig. (4). Here, 
\underline{SDSS}, \underline{MAP}, \underline{P} have the same meaning as 
Fig. in (4b). 
\underline{SNAP} means constraints of the SNAP satellite \cite{SNAP:2001}, 
\underline{current SN} means present constraints using about 50 SNIa, 
\underline{PLANCK} means constraints of the PLANCK satellite 
\cite{PLANCK:2001}.}
\end{center}
\end{figure}

Let us also note that the use of cluster evolution for constraining 
cosmological parameters is not a new idea; in particular, in 1992 Oukbir \& 
Blanchard \cite{oukbir:1992} introduced it and more recently in 2001, 
Sadat \& Blanchard \cite{sadat:2001}, using it again, find a value of 
$\Omega_m$ much higher than $1/3$.
\\
Of course, the three combined measurements will first confirm or infirm the 
existence of dark energy and then -in the case of a confirmation- will give 
information on its nature by measuring or constraining its equation of state.
\section{Conclusion}
We have seen that:
\begin{itemize}
\item $i)$ The accurate determination of the primeval deuterium abundance pins down the 
baryon density of the Universe: $\Omega_B h^2 \sim 0.02$. New CMBA data (BOOMERANG, 
MAXIMA, DASI) lead also to $\Omega_B h^2 \sim 0.02$ and can significantly constrain 
neutrino physics if an additional cosmological constrain is imposed. The 
precision of CMBA measurements will be further improved by MAP and 
PLANCK. These new observations will thus offer new opprotunities to detect or 
constrain new neutrino physics in the early Universe.
\item $ii)$ While luminosity-distance measurements of type Ia SN calibrated candles have 
recently shown that our Universe is accelerating now, the resent question is: what is the 
dark energy ?
Particle physics theory proposes a number of alternatives to a non-zero vacuum 
energy/cosmological constant: quintessence in particular.
\\
With future CMBA measurements (MAP, PLANCK) it should be possible to measure a constant 
equation of state of dark energy within a 10-30 \% accuracy. To determine $\omega$ 
to 5 \% and to begin to probe a time-varying equation of state requires also a large 
SNIa survey (such as SNAP) and a count of rich clusters of galaxies or a 
high-redshift survey of galaxies such as DEEP2 \cite{davis:2000}.
\end{itemize}
The research on the nature of dark energy, however, is still in a nascent stage. Over the 
next decade a variety of new strategies and more precise applications of old strategies 
could very well answer this question once and for all. 

{\bf Acknowledgments.}
\\
The authors would like to thank Francesco and Bianca Melchiorri for their 
helpful discussions and continuous encouragements. We would like to thank 
Marco De Petris, Massimo Gervasi and Fernanda Luppinacci for organizing 
a superb meeting at a wonderful location. Part of the work of D. Puy has been supported by the {\it D$^r$ Tomalla} Foundation and the Swiss National 
Science Foundation.


\begin{thebibliography}{}

\bibitem[1]{scott:1999}
Scott D., \texttt{astro-ph/9911325}.

\bibitem[2]{burles:1999}
Burles S., Nollett K., Turner M., \texttt{astro-ph/9903300}.

\bibitem[3]{signore:1999}
Signore, M., Puy, D., \emph{New Ast. Rev.}, \textbf{43}, 185 (1999).

\bibitem[4]{perlmutter:1999}
Perlmutter, S. et al., \emph{ApJ}, \textbf{517}, 565 (1999).
 
\bibitem[5]{riess:1998}
Riess, A., et al., \emph{Astron J.}, \textbf{116}, 1009 (1998).

\bibitem[6]{signore:2001}
Signore, M., Puy, D., \emph{New Ast. Rev.}, \textbf{45}, 409 (2001).

\bibitem[7]{SNAP:2001}
SNAP, \texttt{http://snap.lbl.gov}.

\bibitem[8]{burles:2001}
Burles S., Nollett K., Turner M., \emph{ApJ}, \textbf{552}, L1 (2001); 
\texttt{astro-ph/0010171}.\\
O'Meara J., Tytler D., Kirman D. et al., \texttt{astro-ph/0011179}.

\bibitem[9]{debernardis:2000}
De Bernardis P., Ade P., Bock J. et al., \emph{Nature}, \textbf{404}, 955 
(2000).

\bibitem[10]{hanany:2000}
Hanany S., Ade P., Balbi A.  et al., \emph{ApJ Lett.}, \textbf{545}, L5 (2000).

\bibitem[11]{lesgourgues:2000}
Lesgourgues J., Peloso M., \emph{Phys. Rev. D}, \textbf{62}, 1301 (2000); 
\texttt{astro-ph/0004412}.

\bibitem[12]{hannestad:2000}
Hannestad S., \emph{Phys. Rev. Lett.}, \textbf{85}, 4203 (2000); 
\texttt{astro-ph/0005018}.

\bibitem[13]{orito:2000}
Orito N., Kajino T., Mathews G., Boyd R., \texttt{astro-ph/0005446}.

\bibitem[14]{esposito:2000}
Esposito S., Mangano G., Melchiorri A. et al., \emph{Phys. Rev. D}, 
\textbf{63}, 3004 (2000); \texttt{astro-ph/0007419}.

\bibitem[15]{mangano:2001}
Mangano G., Melchiorri A., Pisanti O., \emph{Nucl. Phys. Proc. Supp.}, 
\textbf{100}, 369 (2001); \texttt{astro-ph/0012291}.

\bibitem[16]{netterfield:2001}
Netterfield C., Ade P., Bock J. et al., \texttt{astro-ph/0104460}.

\bibitem[17]{lee:2001}
Lee A., Ade P., Balbi A. et al., \texttt{astro-ph/0104459}.

\bibitem[18]{halverson:2001}
Halverson N., Leitch E., Pryke C. et al., \texttt{astro-ph/0104489}.

\bibitem[19]{kneller:2001}
Kneller J., Sherrer R., Steigman G., Walker T., \texttt{astro-ph/0101386}.

\bibitem[20]{hannestad:2001}
Hannestad S., \texttt{astro-ph/0105220}.

\bibitem[21]{hansen:2001}
Hansen S., Mangano G., Melchiorri A. et al. \texttt{astro-ph/0105385}.

\bibitem[22]{steigman:2001}
Steigman G., \texttt{astro-ph/0108148}.

\bibitem[23]{lisi:1999}
Lisi E., Sarkar S., Villante F., \emph{Phys. Rev. D} \textbf{59}, 3520 (1999).

\bibitem[24]{seljak:1996}
Seljak U., Zaldarriaga M., \emph{ApJ} \textbf{469}, 437 (1996).

\bibitem[25]{jaffe:2000}
Jaffe A., Ade P., Balbi A. et al., \texttt{astro-ph/0007333}.

\bibitem[26]{bahcall:1998}
Bahcall N., Fan X., \emph{Proc. Nat. Acad. Sc.} \textbf{95}, 5956 (1998); 
\emph{ApJ} \textbf{504}, 1 (1998).

\bibitem[27]{huey:1999}
Huey G., Wang L., Dave R. et al., \emph{Phys. Rev. D} \textbf{59}, 3005 
(1999); \texttt{astro-ph/9804285}.
 
\bibitem[28]{weller:2001a}
Weller J., Albrecht A., \texttt{astro-ph/0106079}. 

\bibitem[29]{steinhardt:1999}
Steinhardt  P.,Wang I.,Zlatev I., \emph{Phys. Rev. D} \textbf{59},3504 (1999).

\bibitem[30]{zlatev:1999}
Zlatev I., Wang I., Steinhardt P., \emph{Phys. Rev. Lett.} \textbf{82}, 896 
(1999).

\bibitem[31]{ratra:1988}
Ratra B., Peebles P., \emph{Phys. Rev D} \textbf{37}, 3406 (1988).

\bibitem[32]{binetruy:1999}
Binetruy P., \emph{ Phys. Rev D} \textbf{60}, 3502 (1999).

\bibitem[33]{brax:1999}
Brax Ph., Martin J., \emph{Phys. Lett B} \textbf{468}, 40 (1999).

\bibitem[34]{kolda:1999}
Kolda C., Lyth D., \emph{Phys. Lett. B}  \textbf{458} 197 (1999).

\bibitem[35]{weller:2001b}
Weller J., Albrecht A., \emph{Phys. Rev. Lett.} \textbf{86}, 1939 (2001); 
\texttt{astro-ph/0008314}. 

\bibitem[36]{maor:2001}
Maor I., Brustein R., Steinhardt P., \emph{Phys. Rev. Lett.} \textbf{86}, 
6 (2001). 

\bibitem[37]{hu:1999}
Hu W., Eisenstein D., Tegmark M., White M., \emph{Phys. Rev. D} \textbf{59}, 
3512 (1999); \texttt{astro-ph/9806362}. 

\bibitem[38]{SDSS:2001}
SDSS (2001), \texttt{http://www.astro.princeton.edu/BBook}

\bibitem[39]{MAP:2001}
MAP (2001), \texttt{http://map.gsfc.nasa.gov}

\bibitem[40]{huterer:1999}
Huterer D., Turner M., \emph{Phys. Rev. D} \textbf{60}, 1301 (1999); 
\texttt{astro-ph/0012510}. 

\bibitem[41]{PLANCK:2001}
PLANCK (2001), \texttt{http://astro.estec.esa.nl/Planck/}

\bibitem[42]{oukbir:1992}
Oukbir J., Blanchard A., \emph{Astron. Astroph.} \textbf{262}, L21 (1992).

\bibitem[43]{sadat:2001}
Sadat R., Blanchard A., \emph{Astron. Astroph.} \textbf{371}, 195 (2001).

\bibitem[44]{davis:2000}
Davis M., Newman J., Faber S., Phillips A., \texttt{astro-ph/0012189}; 
\texttt{http://astro.berkeley.edu/deep/}.

\end{thebibliography}
\end{document}